\documentclass[10pt,journal]{IEEEtran}
\usepackage[T1]{fontenc}
\usepackage{graphicx,booktabs,array,listings,xcolor,url}
\usepackage[colorlinks,urlcolor=blue,linkcolor=blue,citecolor=blue]{hyperref}
\expandafter\def\expandafter\UrlBreaks\expandafter{\UrlBreaks\do\/\do\*\do\-\do\~\do\'\do\"\do\-}
\definecolor{lgray}{gray}{0.5}
\begin{document}

\title{terms.txt: A Consent and Compensation Protocol for Agentic Web Access}

\author{Rajarshi~Chowdhury%
\thanks{R. Chowdhury is an independent researcher (ORCID 0009-0007-7032-2450). Contact: mailtorajarshi@gmail.com. This work was carried out in the author's personal capacity and does not represent the views of the author's employer.}%
\thanks{Preprint. Submitted to \emph{IEEE Internet Computing}, Special Issue on Future Internet Systems with LLMs and Agents, September 2026. Code and raw results: \url{https://doi.org/10.5281/zenodo.22647915}.}}

\markboth{Preprint, September 2026}{Chowdhury: terms.txt: A Consent and Compensation Protocol for Agentic Web Access}

\maketitle

\begin{abstract}
The open web ran on an unwritten bargain: sites admitted crawlers, and search engines sent visitors back. Public measurements show that bargain breaking under AI crawlers and agents. Automated clients now make up most requests, training dominates Cloudflare-classified crawling, and the largest AI platforms fetch thousands of pages for each visitor they return. The web's common control, \texttt{robots.txt}, cannot express identity, purpose, terms, or price, can be circumvented, and newer alternatives are largely proprietary CDN features. We specify \texttt{terms.txt}, a \texttt{robots.txt}-style file for per-path, per-purpose machine-access terms, plus an origin-enforced exchange using Web Bot Auth signatures, signed intent, delegation tokens, HTTP 402 negotiation, and signed receipts. We define what the exchange can enforce, audit, and leave to contract. A dependency-free implementation adds 0.20 to 0.65 ms per request on one vCPU.
\end{abstract}


\IEEEPARstart{O}{n} September 15, 2026, Cloudflare will change the default terms a new website presents to machines. For every domain onboarded from that day, crawlers in its Training and Agent categories will be blocked on pages carrying advertising unless the owner opts out, and a multi-purpose crawler will be judged by its most restricted purpose \cite{r15}. With more than 20 percent of the web behind Cloudflare \cite{r1}, one vendor's default will set the opening terms of machine access for a large share of new origins.

That one private company can set these defaults, and that publishers welcomed them, shows how badly existing mechanisms have failed. Those mechanisms are the Robots Exclusion Protocol \cite{r8} and the unwritten bargain built around it: sites let crawlers read their pages, operators built indexes, and those indexes sent people back to see ads, click affiliate links, and buy subscriptions.

This article makes three contributions. First, it assembles public measurements from traffic operators and independent researchers to show that the bargain broke, and how quickly. Second, it shows that \texttt{robots.txt} cannot express what a site now needs to tell a machine client and is not reliably honored. Third, it specifies \texttt{terms.txt}, a \texttt{robots.txt}-style file of access terms, and an origin exchange that enforces them. The design composes existing standards work into one request-response flow, runs as origin middleware without a proxy, and measures the cost. It also separates what can be enforced before delivery, audited afterward, and left to contract. The measurements motivate the work; the protocol is the contribution.

\section{The Bargain in Numbers}

The economics of the open web were never written into a protocol. They emerged from HTTP, which serves a page to whoever asks; \texttt{robots.txt}, which lets a site ask clients not to fetch certain paths; and search engines, which turned crawled pages into referrals. Publishers tolerated crawlers because traffic came back.

Three measurement programs show whether it still does, but they count different things. Imperva, analyzing all requests across its customer base, including API calls, put automated clients at 51 percent of web traffic in 2024 and above 53 percent in 2025 \cite{r2}. Cloudflare, counting only HTML page requests behind its network, reported 57.5 percent automated in June 2026 \cite{r1}. \cite{r3}, The figures are not directly comparable, and neither isolates agentic AI traffic. They do point in the same direction.

The metric that captures what comes back is Cloudflare Radar's crawl-to-refer ratio, published since July 2025: HTML pages a platform's crawlers request for each HTML page visit it refers to a site \cite{r4}. Conventional search is near five to one \cite{r3}. From June 19 to 26, 2025, Anthropic's crawlers requested roughly 70,900 pages per referred visit \cite{r4}. Cloudflare later reported Anthropic at 286,000 in January 2025 and 38,000 in July, an 87 percent decline that still left it the most crawl-heavy platform; Perplexity was at 194 and OpenAI near 1,100 \cite{r5}. By mid-2026, Radar's trailing 28-day figure for Anthropic had fallen from about 4,600 in late June to below 2,000 in July as the operator split traffic into purpose-specific agents. OpenAI was in the high hundreds, Perplexity near 190, and Google near five \cite{r3}. Even after the largest improvement in the dataset, the most crawl-heavy AI platform still fetched hundreds of times more pages per visitor than search.

Three caveats matter. Ratios are window-specific and cannot be averaged. Referrals from native applications often lack a \texttt{Referer} header, which can overstate the imbalance \cite{r4}. And the operator publishing the metric also sells the remedy.

Purpose helps explain the ratio because model-training fetches produce no referral by design. Among AI-specific crawlers Cloudflare classifies, training rose from 72 percent of requests in July 2024 to 79 percent in July 2025, while search fell from 26 to 17 percent \cite{r5}. In the broader 2026 basket, which adds a mixed-use category, training was 52 percent of classified crawler requests in June 2026, up from 22 percent in spring 2025, while mixed-use crawlers exceeded 36 percent \cite{r1}. The snapshots use different denominators but point the same way. Mixed use makes the incentive problem structural: Google's crawler serves both search and AI products, so a site that wants visibility in the engine supplying roughly 88 percent of referral traffic cannot refuse that crawler's AI use. Cloudflare estimates this gives Google about twice the information access of leading AI companies \cite{r1}.

\section{Where the Clicks Went}

In March 2025, Pew Research Center recorded the browsing of 900 U.S. adults and captured 68,879 Google searches. When an AI-generated summary appeared, users clicked a traditional result in 8 percent of visits versus 15 percent without one, clicked a source cited in the summary in 1 percent, and ended the session in 26 percent versus 16 \cite{r6}. Pew does not claim causation. It also reconstructed AI-summary exposure by re-running recorded queries rather than capturing what each panelist saw. The crawl still happens. The click often does not.

\section{A Text File From 1994}

Against this backdrop, most sites still rely on \texttt{/robots.txt}. Proposed in 1994 and codified as RFC 9309 in 2022, it tells a conforming crawler which rules to follow but provides no server-side enforcement and is not access control \cite{r8}. Its vocabulary is allow or disallow, by named user agent and path prefix. It cannot distinguish training from indexing or a live answer request, state a price, license, or rate, or verify that a client calling itself \texttt{GPTBot} really is \texttt{GPTBot}. Later files in the same mold, including \texttt{humans.txt}, \texttt{security.txt}, \texttt{ads.txt}, \texttt{llms.txt} \cite{r20}, and \texttt{ai.txt} \cite{r19}, are also declarations with nothing behind them; \texttt{ai.txt} says so explicitly. \texttt{robots.txt} is a preference signal being asked to do the work of a licensing system, and it fails in four ways.

First, it produces incoherent policy. Longpre and colleagues audited 14,000 domains underlying the C4, RefinedWeb, and Dolma corpora \cite{r7}. From April 2023 to April 2024, \texttt{robots.txt} restrictions on AI crawlers went from nearly nonexistent to covering more than 5 percent of tokens in those corpora and more than a quarter of tokens from the most actively maintained domains. Terms-of-service restrictions covered 45 percent of C4, and the two channels often contradicted each other. The authors warn that because the file cannot distinguish a commercial training crawler from an archive or research crawler, publishers reacting to the former also block the latter.

Second, the web cannot agree on what to say. Radar's directive analysis finds \texttt{GPTBot} is both the most explicitly disallowed AI crawler and the most explicitly allowed \cite{r3}, a sign that the format is standing in for a negotiation it cannot support.

Third, it is not reliably honored. In August 2025, Cloudflare documented Perplexity fetching pages from sites that had disallowed its declared crawler, using undeclared user agents and rotating source networks \cite{r9}. Whatever one concludes about that dispute, the core weakness is clear: the mechanism relies on the client truthfully identifying itself, with no way to verify the claim.

Fourth, it is blunt where precision matters. With more than a third of classified crawler requests coming from mixed-use bots \cite{r1}, a rule meant to block training can also remove a site from discovery. Cloudflare frames the choice for a small site as allowing AI training or losing discoverability \cite{r15}.

\section{What Exists Today, and the Gap}

A replacement is taking shape across four threads. Table 1 compares them.

For identity, Web Bot Auth lets an automated client sign requests with a key published in a well-known directory on its operator's origin, allowing a server to verify the claimed operator. It builds on RFC 9421 HTTP Message Signatures \cite{r13}, carries the operator's HTTPS origin in a \texttt{Signature-Agent} dictionary keyed by signature label, and identifies keys by JWK thumbprint. On September 1, 2026, the IETF Web Bot Auth working group adopted it as a Standards Track document \cite{r11}. A companion draft defines a Signature Agent Card that advertises identity, purpose, and rate expectations \cite{r12}. The draft explicitly excludes authorization, delegation, and an intent vocabulary.

For preferences, the IETF AI Preferences working group is standardizing a vocabulary for how automated systems may use content, plus a way to attach those preferences through \texttt{robots.txt} and HTTP headers. The vocabulary is at revision 07, and the attachment draft, revised in August 2026, updates RFC 9309 \cite{r10}. The charter excludes enforcement and authentication, so the result is intentionally a richer preference layer, not access control.

For pricing, Cloudflare's Pay Per Crawl, in private beta since July 1, 2025, lets a site allow, charge, or block each identified AI crawler. Charged crawlers receive HTTP 402 Payment Required, with Cloudflare as merchant of record \cite{r14}. On July 1, 2026, Cloudflare said it was beginning to reshape this into Pay Per Use, paying publishers when content appears in an answer rather than per fetch because fetch counts are a poor proxy for value. It calls this an experiment with Ceramic.ai and You.com \cite{r16}. Independent gateways use a similar model: Fairfetch returns 402 with a price and usage category and settles over x402, but the agent fetches through Fairfetch's endpoint rather than the origin \cite{r18}.

Cloudflare currently ships the most complete proprietary composition. Its July 2026 taxonomy separates Search, Agent, and Training crawlers. A content-use signal, \texttt{use=immediate}, reference, or full, extends the Content Signals convention in \texttt{robots.txt} but remains a preference. A Forwarded header carrying \texttt{for="openai"} with \texttt{use="reference"} conveys transitive trust through intermediaries using RFC 7239. Verified status is no longer default-allow, can be revoked when a bot abuses content-use signals, and is unavailable to bots that reproduce content in full \cite{r15}.

\begin{table*}
\caption{Mechanisms governing machine access and what each expresses and enforces, as of September 2026.}
\label{tab1}
\footnotesize
\begin{tabular*}{\textwidth}{@{}p{0.149\textwidth}<{\raggedright}p{0.105\textwidth}<{\raggedright}p{0.131\textwidth}<{\raggedright}p{0.122\textwidth}<{\raggedright}p{0.105\textwidth}<{\raggedright}p{0.114\textwidth}<{\raggedright}p{0.114\textwidth}<{\raggedright}@{}}
\toprule
\textbf{Mechanism} & \textbf{Identity verified} & \textbf{Purpose} & \textbf{Terms or price} & \textbf{Delegation} & \textbf{Enforced where} & \textbf{Status} \\
\midrule
\texttt{robots.txt} (RFC 9309) & No & No & No & No & Nowhere, advisory & Standard \\
\texttt{llms.txt}, \texttt{ai.txt} & No & Partly (\texttt{ai.txt} actions) & No & No & Nowhere, preference & Community proposals \\
AIPREF vocab and attach & No & Yes (\texttt{train-ai}, search) & No & No & Nowhere, by charter & IETF WG drafts 07 and 05 \\
Content Signals use= & No & Partly (use level) & No & No & Nowhere, preference & Vendor convention \\
Web Bot Auth & Yes (RFC 9421) & Per bot, via Agent Card & No & No & Origin or proxy & IETF WG document, Sept. 2026 \\
Forwarded transitive trust & Relies on WBA & Per hop & No & Operator only & Proxy & Vendor proposal \\
Pay Per Crawl (402) & Vendor bot directory & Per bot category & Price per fetch & No & Proxy only & Vendor beta, Pay Per Use experiment announced \\
Fairfetch gateway (402, x402) & No & Usage category & Price per fetch & No & Gateway only & Open-source product \\
\texttt{terms.txt} exchange & Yes (Web Bot Auth) & Per request, signed & \texttt{terms.txt}, per path and purpose & Yes, agent-bound, scoped & Any origin & This article \\
\bottomrule
\end{tabular*}
\end{table*}

The open, multi-vendor, standards-track work verifies identity but not terms. The preference work can express intended use but, by charter, does not enforce it. Every mechanism that does enforce access today is tied to a proxy, so a site gets it only by routing traffic through a particular company. A publisher on Cloudflare's free plan gets a purpose-aware allow list and a 402 negotiation surface; a publisher running its own server gets a text file. Meanwhile, the unit of account moved from per crawl to per use within a year, showing that the market has not settled what it is pricing. The rest of this article shows that access terms and compensation need not live in a vendor dashboard.

\section{Design: A Consent and Compensation Exchange}

We compose the existing pieces into one exchange at the HTTP request boundary. That is the one point where identity, purpose, and terms meet, and it lets an origin enforce policy without renting a proxy. Figure 1 shows the architecture and flow.

\begin{figure*}
\centerline{\includegraphics[width=\textwidth]{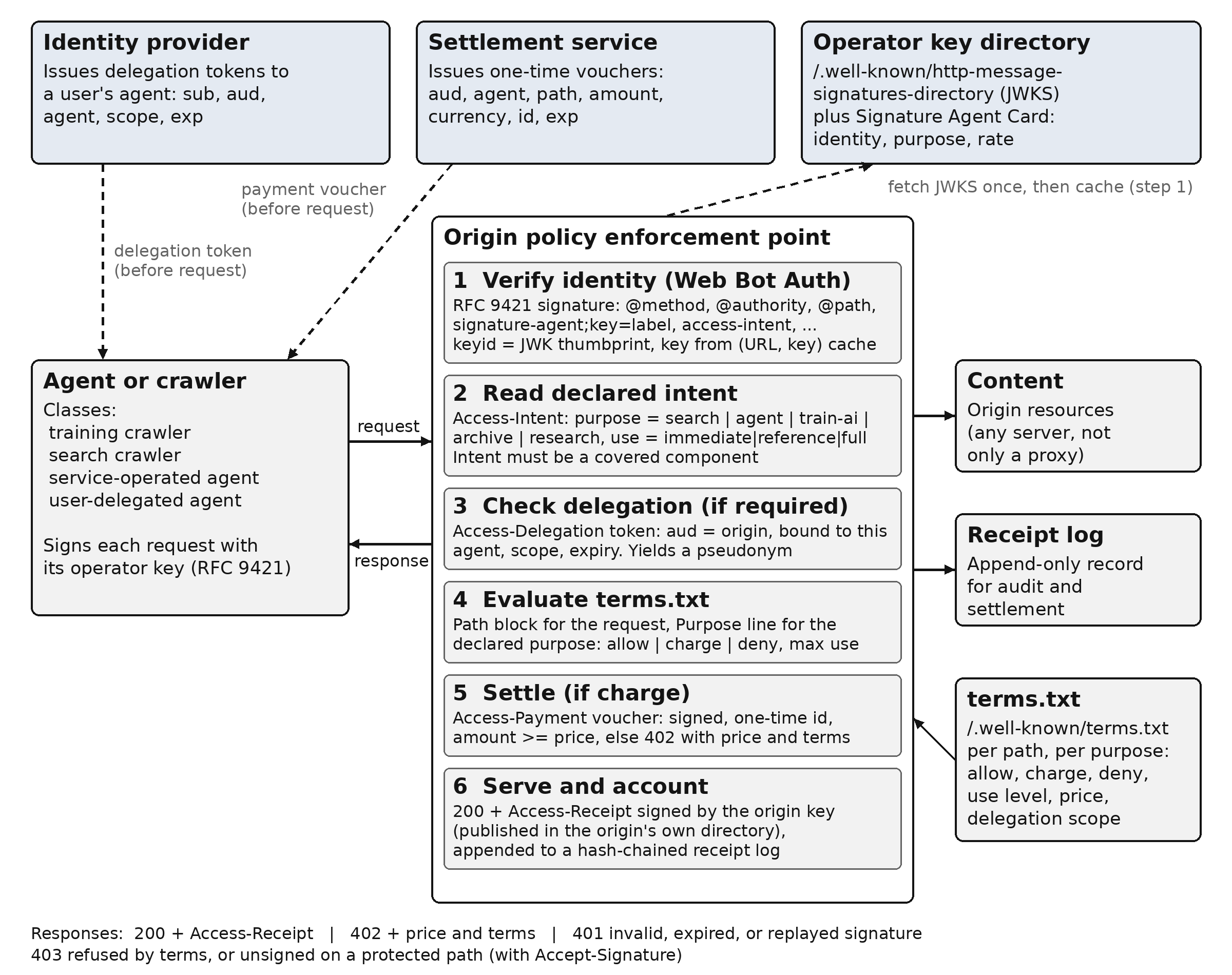}}
\caption{The consent and compensation exchange at the HTTP boundary. Solid arrows show the request path; dashed arrows show out-of-band setup completed beforehand.}
\label{fig1}
\end{figure*}

Because terms differ by client, the design separates four client classes. A training crawler fetches content to build a model. A search crawler fetches it to build an index that refers traffic. A service-operated agent fetches in real time for a service, such as an answer engine grounding a response. A user-delegated agent fetches for a specific person and should inherit that person's entitlements, such as a subscription, without revealing that person's identity to the origin.

In the exchange, an agent holds its operator's signing key and, when needed, a delegation token from the user's identity provider and a payment voucher from a settlement service. All are obtained before the request. The agent then sends an ordinary HTTP request with up to six headers. \texttt{Signature-Agent} uses the working-group draft's dictionary form, \texttt{sig1="https://bot.example"}, to name the operator's origin; its well-known key directory is fetched once and cached under the (URL, key) pair. \texttt{Signature-Input} and Signature carry an RFC 9421 signature over \texttt{@method}, \texttt{@authority}, \texttt{@path}, the signature-agent member keyed to the label, \texttt{Access-Intent}, and, when present, \texttt{Access-Delegation} and \texttt{Access-Payment}. Validity is limited to five minutes and includes a nonce, a \texttt{keyid} containing the JWK SHA-256 thumbprint, and \texttt{tag="web-bot-auth"}. \texttt{Access-Intent} is an RFC 9651 dictionary declaring purpose and use, for example \texttt{purpose="agent", use="reference"}. \texttt{Access-Delegation} and \texttt{Access-Payment} carry compact signed tokens. The origin verifies the signature, atomically reserves the nonce before any discovery fetch, parses intent, checks required delegation and scope, evaluates the path-and-purpose terms, and verifies payment when required. It then serves the content with a signed \texttt{Access-Receipt}, returns 402 with price and terms, 401 for identity failures, or 403 for policy refusals. Unsigned requests are handled per path: public paths are served, while protected paths return 403 with an \texttt{Accept-Signature} challenge.

\noindent\begin{minipage}{\columnwidth}
\texttt{terms.txt} is the discoverable half. An origin publishes \texttt{/.well-known/terms.txt}, a \texttt{robots.txt}-style file that any client can fetch before its first request:
\begin{lstlisting}
# /.well-known/terms.txt
Terms-Id: 2026-09-01
Receipt-Keys: https://origin.example/.well-known/http-message-signatures-directory
Payment: voucher https://settle.example/.well-known/settlement

Path: /articles/
Unsigned: allow
Purpose: search    allow   use=reference
Purpose: agent     allow   use=reference  delegation=read:articles
Purpose: train-ai  charge  0.002 USD/request
Purpose: archive   allow   use=full

Path: /premium/
Unsigned: challenge
Purpose: agent     charge  0.01 USD/request  use=reference  delegation=read:premium
Purpose: train-ai  deny
\end{lstlisting}
\end{minipage}

The header fields name the terms version, the origin's key directory, and accepted payment methods. The receipt-signing key is published in that directory under its thumbprint so anyone can verify receipts. Each Path block sets the rule for unsigned requests and, for each purpose, an allow, charge, or deny decision with optional use ceiling, price, and delegation scope. The purpose vocabulary, search, agent, \texttt{train-ai}, archive, and research, maps to AIPREF terms and Cloudflare's taxonomy while adding the two categories that the Longpre audit shows are caught in the crossfire. The use levels match Content Signals' three levels, so a site's \texttt{robots.txt} preferences and enforceable terms can use the same language. As with \texttt{robots.txt}, the file itself enforces nothing; enforcement comes from the paired exchange.

Delegation is agent-bound and pseudonymous. A delegation token binds a pairwise pseudonymous subject to a specific operator, audience, named scope such as \texttt{read:premium}, and expiry. A stolen token is therefore useless to another operator, and one entitlement does not unlock another. Because the subject is a pairwise pseudonym from the identity provider, the origin learns that a subscriber's agent is present without learning the subscriber's identity. Vouchers are bound the same way and include a one-time identifier.

Receipts are the accounting primitive. For every served request, the origin returns an \texttt{Access-Receipt} signed with its own key. It contains the operator, any subject pseudonym, declared purpose and use, terms identifier, path, a hash of the request signature, a timestamp, and an identifier. The origin appends each receipt to a hash-chained, append-only log, whose head commits to every receipt issued. This gives both sides a non-repudiable record of what was delivered under which terms, regardless of the unit a settlement model pays by. The protocol does not choose the unit of account; it makes the unit measurable.

For caching and intermediaries, receipted responses are marked private and vary on \texttt{Signature-Agent}, \texttt{Access-Intent}, and \texttt{Access-Delegation}, preventing a shared cache from giving one operator another's receipt. Unsigned responses remain cacheable as they are today. A CDN can forward the request unchanged for the origin to verify or verify it itself and pass a Forwarded assertion, as Cloudflare proposes. If an intermediary rewrites a covered component, the signature fails, which is the intended behavior.

\section{What Is Enforced, What Is Audited, What Is Contractual}

A signature over a declared purpose proves who made the declaration and that it was not altered; it does not prove the declaration is true. A signed receipt proves content was delivered under stated terms; it does not prove what happened to the bytes afterward. Once content leaves the origin, HTTP cannot govern its use. The design should state that boundary plainly; much product literature does not.

Before delivery, the exchange enforces several facts: the request came from the named operator, is fresh, and is not a replay; the intent was signed rather than inserted by an intermediary; any delegation is valid, in scope, and bound to that operator; the terms permit that purpose and use level on that path; and a charged request carries valid, unspent payment. The prototype refuses requests that fail any of these checks. Its test suite also covers three bypasses an earlier version allowed: a client-supplied mode header, unsigned access to a protected path, and two identical signed requests racing a cold key lookup.

After delivery, the declared purpose is auditable against behavior. An operator that declares \texttt{use="reference"} but reproduces content in full, or declares search but never refers traffic, leaves evidence in receipts, citations, and traffic. The signed declaration makes that behavior attributable. The remedy is revocation of standing, as Cloudflare now does with Verified status \cite{r15}.

What remains contractual is what a model does with content it lawfully received, including whether it trains on it. A protocol can make terms explicit, attributable, and priced, but it cannot make them self-executing. A header layer does not prevent training, and this design does not assume that it can.

In the threat model, false purpose claims cannot be prevented; they are handled through audit and revocation. Credential theft is limited by validity windows of minutes, key rotation through the directory, and agent binding of tokens and vouchers, so a stolen token cannot be spent by another operator. Replay is blocked by an atomic check-and-reserve on (operator URL, \texttt{keyid}, nonce) before any asynchronous discovery, matching the draft's deployment guidance and fixing an ordering bug in the prototype's first version. Rewriting a covered component at a proxy breaks verification. The origin learns only the operator and a pairwise pseudonym, and receipts carry no direct user identifier. This layer does not stop unauthenticated scraping because it cannot force a client to sign. Instead, it makes honest access more capable: only signed requests can receive delegated entitlements, paid content, or receipts, and sites can require signatures on costly paths. Cost-based denial of service is the main systems risk because verification consumes CPU on every request. Mitigations are to check expiry, nonce, and key presence before cryptography, rate-limit by operator URL, and exploit the measured fact that refusal costs less than service.

\section{Reference Implementation and Per-Request Cost}

We implemented \texttt{terms.txt} and the exchange in about 600 lines of dependency-free JavaScript on Node.js 22: a signature-base library with the \texttt{terms.txt} parser, an origin enforcement point, a directory resolver, and a harness. The library verifies the working-group draft's Ed25519 test vector E.2.1 and reproduces its thumbprint. The draft's printed signature base omits quotes around the \texttt{Signature-Agent} member, but the vector verifies only with them, as RFC 9421 requires. Discovery is bounded by size, key count, and time, refuses redirects, and coalesces concurrent fetches. The harness runs 24 checks each time. In addition to the identity and policy cases above, it verifies that \texttt{terms.txt} is served as text and parses to the enforced policy; key lookup is scoped to the (URL, key) pair; bad or cross-agent delegation is rejected; public unsigned requests are served and protected ones challenged; a client-supplied mode header is ignored; two identical signed requests racing a cold key lookup yield exactly one success; and the receipt log's hash chain matches the reported count and head. Benchmark mode is process configuration, with baseline and identity-only servers running as separate processes.

Table 2 reports loopback end-to-end overhead against the passthrough server. One 2.1 GHz Xeon vCPU was shared by the load generator and all server processes, using loopback without TLS. We ran five independent tests at concurrency 1 and five at 32, each with fresh processes and 10,000 measured requests per scenario after 3,000 warm-up requests. Headers were pre-signed, so client signing is excluded, while the load generator's round trip is included. The passthrough control ran first and last in every run. Unqueued latency was identical in both positions, but first-position throughput was about half the final value because of warm-up, so the table uses the final control as baseline. The key directory was fetched once per agent origin per process, taking about 35 ms on loopback. Across ten runs, median Ed25519 verification was 120 microseconds (118 to 122), delegation verification 122, receipt signing 42, and signature-base construction 4.

\begin{table*}
\caption{Loopback end-to-end overhead relative to the passthrough server: median over five independent runs, with ranges in brackets. One shared vCPU, no TLS.}
\label{tab2}
\footnotesize
\begin{tabular*}{\textwidth}{@{}p{0.239\textwidth}<{\raggedright}p{0.124\textwidth}<{\raggedright}p{0.138\textwidth}<{\raggedright}p{0.086\textwidth}<{\raggedright}p{0.167\textwidth}<{\raggedright}p{0.086\textwidth}<{\raggedright}@{}}
\toprule
\textbf{Scenario} & \textbf{Result} & \textbf{p50 ms at conc. 1} & \textbf{Added ms} & \textbf{Req/s at conc. 32} & \textbf{Header bytes} \\
\midrule
Passthrough control (final) & 200 & 0.054 [0.052-0.056] & 0 & 15,892 [15,079-16,604] & 0 \\
Unsigned, allow path & 200, no receipt & 0.069 [0.066-0.076] & 0.015 & 8,920 [8,529-9,782] & 0 \\
Identity only, three Web Bot Auth headers & 200 & 0.255 [0.246-0.262] & 0.201 & 3,027 [2,856-3,152] & 392 \\
Signed search, receipt and log & 200 + receipt & 0.392 [0.386-0.425] & 0.339 & 2,424 [2,311-2,485] & 458 \\
Delegated agent, receipt and log & 200 + receipt & 0.548 [0.534-0.563] & 0.494 & 1,846 [1,765-1,873] & 825 \\
Charged path, no payment & 402 + terms & 0.426 [0.418-0.431] & 0.373 & 2,227 [2,152-2,254] & 825 \\
Charged path, voucher, receipt and log & 200 + receipt & 0.704 [0.691-0.729] & 0.650 & 1,438 [1,395-1,459] & 1,206 \\
Forged signature & 401 & 0.266 [0.260-0.290] & 0.212 & 3,554 [3,390-3,608] & 458 \\
\bottomrule
\end{tabular*}
\end{table*}

Three results stand out. Identity verification is the largest single cost; intent, \texttt{terms.txt} evaluation, receipt, and logging add about 0.14 ms beyond it. Refusal is cheaper than service: a 402 adds 0.37 ms versus 0.65 for the paid path, and a forged signature is rejected in 0.21 ms, limiting the leverage of a cost-based attack. Throughput falls from about 15,900 to 1,400--3,000 requests per second because one shared core runs Ed25519 in a single thread. Worker threads or a native backend would raise that ceiling, while 2,400 requests per second per vCPU for the common search path already exceeds most origins' load. These are marginal loopback costs, not Internet latency; real-traffic deployment remains to be studied. The source, README, MIT license, and ten raw result files are archived at https://doi.org/10.5281/zenodo.22647915 (release v0.1 of https://github.com/rch0wdhury/terms-txt).

\section{Objections}

Referrals are the wrong metric. In part, yes, and that strengthens the case. An agent can compare twenty retailers, buy from one, and deliver a conversion without a referral trail. If value no longer flows mainly through clicks, accounting has to move to the one point all parties share: the request. Cloudflare's shift from per-crawl to per-use pricing reflects the same search for a value-based unit of account.

Pricing access will enclose the open web. The Longpre audit suggests enclosure is already happening, and crudely, because the current instrument cannot distinguish a research crawler from a commercial pipeline \cite{r7}. A protocol that lets a site say yes to archives, researchers, and indexing, but no or pay to commercial training, can reduce that overblocking.

Operators will not comply. Signed requests make noncompliance detectable rather than invisible, and only signed requests receive delegated entitlements, paid content, or receipts. The EU AI Act also requires general-purpose model providers to honor machine-readable reservations of rights, which is easier to demonstrate against one standard signal than a patchwork of proprietary ones \cite{r17}.

\section{What Can Be Done Now, What Must Be Standardized, What Would Change Our Mind}

Any origin can implement the core exchange today without a proxy: publish \texttt{terms.txt}, verify signatures under the Web Bot Auth draft, cover an intent header with the signature, and issue receipts. Our prototype does all four in about 550 lines, and equivalent nginx, Apache, or Caddy modules are a matter of engineering. Adopting sites immediately gain attributable logs and a negotiation surface, even before operators agree to pay.

Several pieces still need standardization. AIPREF's vocabulary should bind to a signed request-side intent component; today it exists only as a response-side preference. Web Bot Auth needs a delegation token format and scoping model, which neither charter covers. \texttt{terms.txt} needs a grammar and well-known location, and settlement systems need a receipt format they can consume. None is large, and the Web Bot Auth group's September 2026 adoption of its protocol draft creates a natural place to raise them.

The argument is falsifiable. If signed per-request intent does not reduce mixed-use crawling once deployed, purpose declaration is not the lever. If sites that publish terms and receipts see no change in operator behavior or crawl-to-refer ratios over a year, the incentive does not live at the origin. If receipt logs fail to reconcile with answer-engine citation reports, receipts are not a usable unit of account. A large publisher could test the first two within a year of deployment.

Validation would look like a public, vendor-neutral dataset of crawls, referrals, and receipts with standard definitions. That would let the next version of this argument rely less on measurements from a party that also sells the remedy. The bargain that financed the open web was never written down, and the numbers say it is gone. Its replacement may live in the network or in a vendor dashboard. The pieces needed to put it in the network now exist.

\section{A Note on the Data}

Cloudflare Radar figures come from the public API: crawl-to-refer at \path{radar/bots/crawlers/summary/crawl_refer_ratio} and purpose at \path{radar/ai/bots/summary/crawl_purpose}. Values were retrieved from June through August 2026 and rounded. Table 2 is computed by \texttt{aggregate.js} from the archived result files.

\end{document}